# Low Field Magnetic Susceptibility and the Hidden Order Transition in $URu_2Si_2$

B.S. Shivaram[1,2] and D. G. Hinks[2]

[1]Department of Physics, University of Virginia, Charlottesville, VA. 22901

[2]Argonne National Labs, Argonne, IL. 60639.

## ABSTRACT

Despite several decades of both experimental and theoretical work the nature of the hidden order transition at $T_{HO}$ = 17.5 K in $URu_2Si_2$ remains enigmatic. We report here low field DC magnetization as well as AC susceptibility measurements performed on single crystals with field along the c-axis. A peak in the first order AC susceptibility is observed at T=16 K in close proximity to the well known change in slope at $T_{HO}$=17.5 K. Third order susceptibility measurements on the other hand reveal a signature only at 16 K with no discernible change at $T_{HO}$. However, both transitions are visible in the fifth order susceptibility. A signature at this lower transition, as a spontaneous ferromagnetic signal, also appears in low field DC magnetization measurements. A similar but larger ferromagnetic response appears at T~35 K. Both these ferromagnetic signatures are <u>suppressed</u> in a small field (B ~ 40 Gauss ||c-axis). The close proximity of the lower signature to the hidden order transition and the simultaneous presence and suppression of a ferromagnetic signature at 35 K could be due to the removal of degeneracy in the complex order-parameter of the hidden order transition due a "symmetry breaking field", at least locally within the crystals of $URu_2Si_2$.

Uranium based heavy electron metals such as $URu_2Si_2$, $UPt_3$, and $UPd_2Al_3$ have been a rich testing ground for many new ideas in condensed matter physics. Considerable progress in heavy fermion physics has been driven by the interest in superconductivity and it's interplay with magnetism[1]. The superconducting order in these materials can be tensorial in character with it's coupling to vector magnetism giving rise to novel effects such as multiple superconducting phases and a complex phase diagram[2,3]. The magnetic properties of these materials are equally intriguing with for example both $UPt_3$ and $URu_2Si_2$ exhibiting transitions at a temperature T ~10 $T_C$ with very small ordered moments of the order of 0.01$\mu_B$. In $UPt_3$ it has been difficult to detect changes in the thermodynamic properties at this temperature a fact consistent with the smallness of the ordered moment. On the other hand, large changes are observed in $URu_2Si_2$, in several of the measured properties at $T_{HO}$=17.5 K. Thus the nature of the order parameter in this material and how the small ordered moment can be reconciled with the observed changes in the thermodynamic properties at $T_{HO}$ are matters of current debate[4]. A number of recent experiments[5] as well as theories[6] have addressed this question. The new experiments have focused mostly on extremely perfect single crystals with no ferromagnetic impurities[7] and superconducting $T_c$'s typically at 1.4 K. In the present work we examine single crystal samples which were intentionally not annealed and where a strong ferromagnetic signature is present and superconducting $T_c$=1.05 K[8]. We find a number of new results which might be equally relevant in solving the puzzle of the HO phase in $URu_2Si_2$.

The measurements we report were performed on four pieces of single crystals obtained from an oriented single crystalline rod grown using vertical float zone refining at Argonne National Labs. A commercial magnetometer (Quantum Design. MPMS ll SQUID magnetometer) was used to obtain the DC magnetization in a straightforward way. However a slight modification of the commercial setup was needed for the AC susceptibility measurements. The linear and nonlinear contributions to the AC magnetic response were obtained by monitoring the fundamental as well as the harmonics generated at the SQUID amplifier in the MPMS II in a manner similar to that employed in earlier work by Levy[9]. A digital lock in amplifier was used to measure the harmonic content of the signals. The phase of the detection circuitry was adjusted utilizing the superconducting transition of a small speck of lead.

In figure 1 we show the real and imaginary parts of the AC response at a fundamental frequency of 3.8 Hz with the excitation field along the c-axis. Clearly noticeable is the well known change in the slope of the linear susceptibility at $T_{HO}$=17.5 K. The change in slope of the linear susceptibility is a characteristic of a second order transition. Also visible clearly is a second signature at a temperature roughly 1 K lower. Further clues to this unusual magnetic response in our crystals of $URu_2Si_2$ are provided in the following results. In figure 2 we show the results obtained at the third harmonic of the excitation frequency. The magnitudes of the in phase and quadrature signals at this harmonic are a measure of the real and imaginary parts of the third order susceptibility respectively. The position of the dissipative peak, shown in the lower part of the figure, corresponds to the signature at the lower temperature shown in figure 1 (rather than to the upper transition at 17.5 K). In fact there is no discernible third order susceptibility change at the upper transition. The AC measurements have also been repeated at a frequency of 0.2 Hz, and we obtain the same results as in figures 1 and 2. Thus eddy current effects which plague measurements at higher frequencies are absent in our case. In figure 4 we show the signals obtained at the next higher odd harmonic for $\chi_5$. While a clear signature in the dissipative part is again visible at the lower transition a signature of comparable magnitude is also seen at the upper transition.

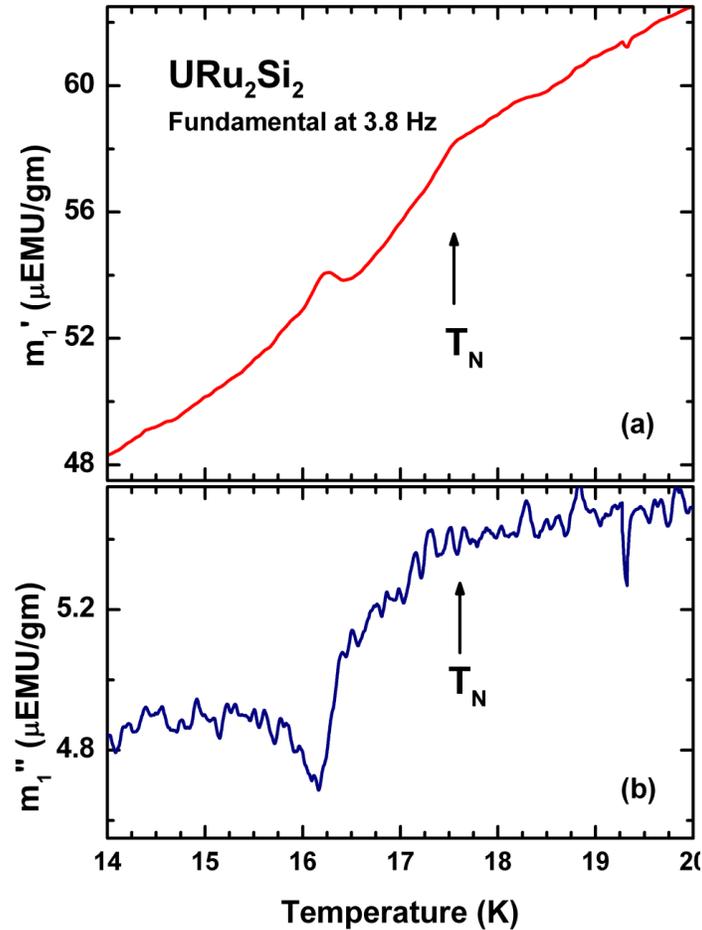

*Figure 1:* Shows the real and imaginary parts of the linear susceptibility in $URu_2Si_2$. Note the feature at 16 K in addition to the well known change in slope at $T_{HO}$=17.5 K in the real part of $X_1$. The imaginary part is approx. 100 times smaller than the real part. The applied AC field is at 3.8 Hz with an amplitude of 4 gauss.

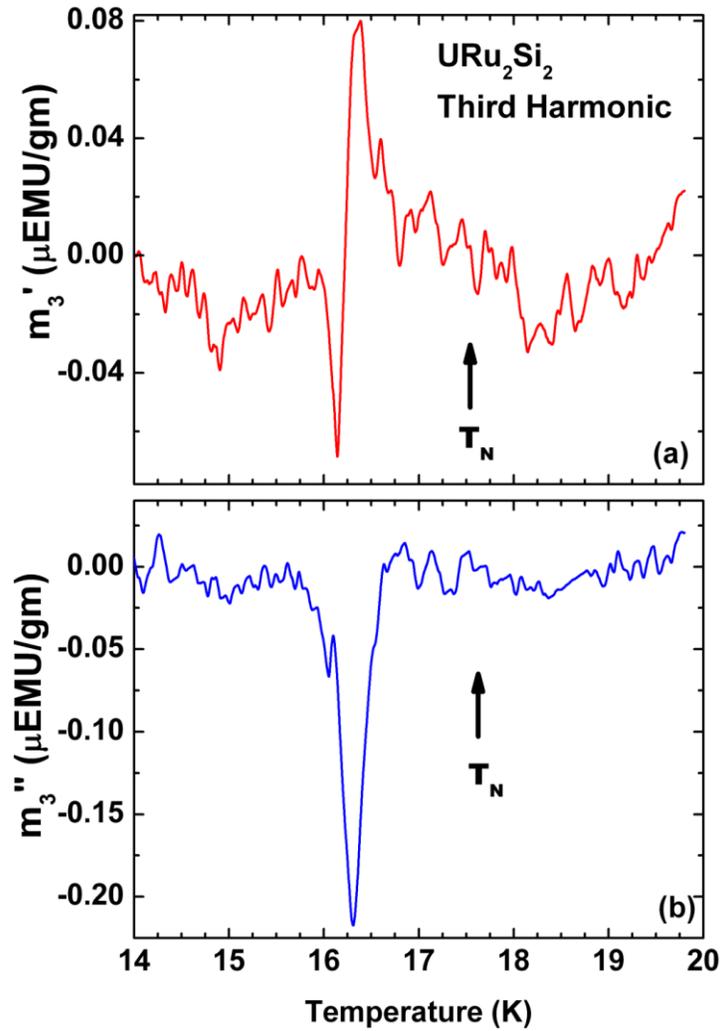

*Figure 2:* Shows the real and imaginary parts of the third order AC susceptibility. Note the absence of a discernible signature at $T_N$ but a strong signature at the new feature at approximately 16 K. The AC applied field is same as in fig. 1.

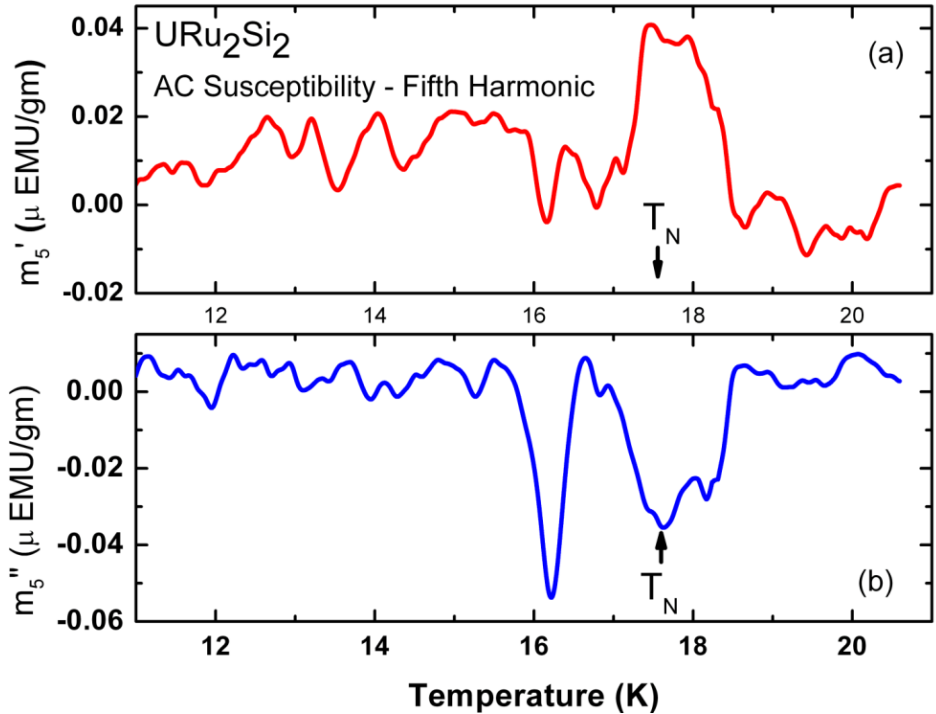

*Figure 3:* Shows the real and imaginary parts of the fifth order AC susceptibility in URu2Si2 in the vicinity of the double transition. The AC applied field is same as in fig. 1.

We have also performed low field DC magnetization measurements on the same samples with B||c-axis. For these measurements the samples were cooled in a field set as close to zero as possible[10] and ramped to successive field strengths. The results obtained are shown in fig.4. A spontaneous ferromagnetic signal arises near 35 K and grows to saturation as the temperature is lowered. But with successive increased applied field the extent of the growth of this signal at low temperatures is <u>reduced</u>. It thus appears that the paramagnetic background grows at the expense of the ferromagnetic behavior. This unusual result originating at T= 35 K has been observed before by Roy et. al.[11] While noting the novelty of this behavior Roy et. al. attribute the effect to the 'faulty' nature of their polycrystalline samples. Our observations, similar in nature, are made on single crystals and could point to a more intrinsic effect.

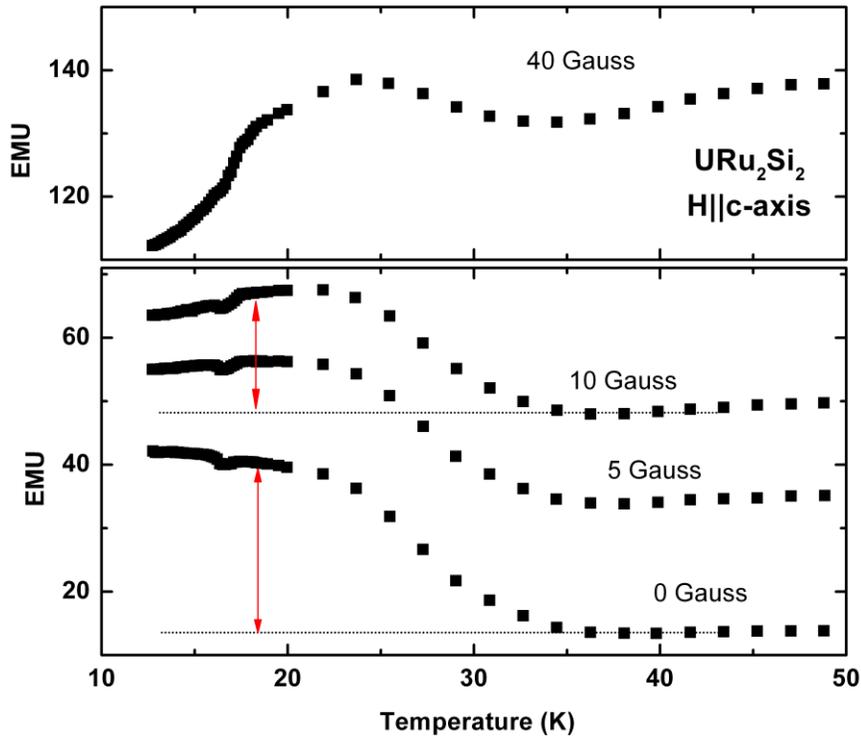

***Figure 4:*** Shows the DC magetization measured in small magnetic fields.  A ferromagnetic component develops at 35 K and grows to saturation at low temperatures.  However, the growth of this component is <u>suppressed</u> by an increasing applied field as indicated by the vertical length of the two red arrows referenced to a high temperature baseline.

     In figure 5 we show the same data as in figure 4 but with the focus on the temperature region near the hidden order transition.  In zero applied field there is an additional growth in the ferromagnetic signal at 16 K (apart from a rapid decrease in the magnetization at 17.5 K) and this growth is also suppressed as the magnetic field is increased.  Since the two ferromagnetic signatures, the one at 35 K and at 16 K behave in a similar manner with applied magnetic field they probably have a common origin.  Further, the paramagnetic background grows at the expense of the ferromagnetic signal suggesting an intimate coupling between two sets of electrons, one which wants to order ferromagnetically and the other which wants to order at the HO transition.

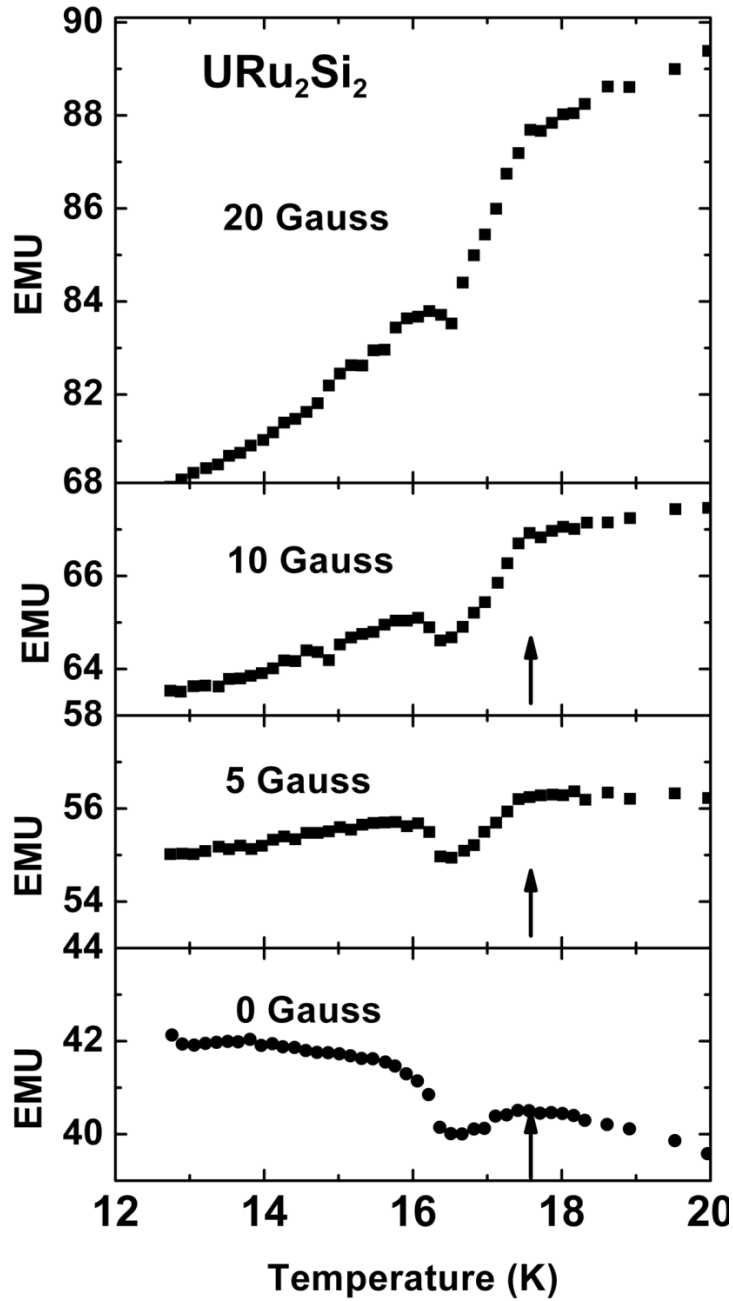

*Figure 5:* Shows the DC magnetization measured in small magnetic fields in the vicinity of the hidden order transition(the position of which is indicated by arrows). A ferromagnetic component which develops at 16 K in zero field is quickly <u>suppressed</u> by an increasing applied field.

We have also performed all of the above discussed measurements with the samples oriented with the basal plane parallel to the longitudinal axis of the SQUID detection coils. Although weak signatures are observed both at the T=35 K ferromagnetic transition and in the vicinity of $T_{HO}$ we are unable to distinguish the splitting for this orientation.

We next turn to a discussion of our results and their relation to results reported in the literature on samples of varying quality. A dependence of the low field DC magnetization on thermal treatment of samples has been noted by Park et. al.[12] In this work a significant difference in the magnetization was obtained between field cooled (FC) and zero field cooled (ZFC) modes for both annealed and unannealed single crystals when the magnetic field B//c-axis. However, the FC magnetization was always larger than the ZFC magnetization. In addition, Park. et. al. found no evidence for a ferromagnetic signature in their samples. Both these results are at variance with the observations of Roy et. al. on polycrystalline samples as well as the DC magnetization results on single crystals obtained by us. On the other hand our results on single crystals are in good agreement with the work of Roy et. al.

Influence of sample quality on the magnetic properties of $URu_2Si_2$ has also been studied by Fak et. al[13] who also report heat capacity and neutron scattering measurements. While their work does not contain any low field magnetic measurements the heat capacity results are noteworthy. While an annealed single crystal exhibits a peak in the heat capacity at the expected temperature of $T_N$=17.5 K, the unannealed crystal has a peak shifted to lower temperature, centered at 16 K. A hint of a possible transition at 16 K in addition to the HO transition at 17.5 is also contained in the NMR measurements of Amitsuka and Yokoyama[14]. The antiferromagnetic volume as deduced from NMR measurements has a change in slope around 16 K at low pressures which disappears at higher pressures as the large moment antiferromagnetic order (LMAF) sets in. We do not know about the existence or otherwise of the ferromagnetic feature at 35 K in the samples used by the investigators in refs. (13) and (14). However, previous investigators performing µSR studies have noted the presence of the 35 K ferromagnetic feature in their samples[15].

**Table I:** Summary of existing results on the saturation magnetization observed in samples with a ferromagnetic feature arising at 35 K.

| REFERENCE | 35 K feature | Saturation Magnetization (EMU/Mole) | Saturation Magnetization ($\mu_B$/U-atom) | 16 K feature |
|---|---|---|---|---|
| Knetsch et. al. [15] | YES | -- | $1.2 \times 10^{-4}$ | -- |
| Roy et. al.[11] | YES | 0.30 | | YES |
| Ramirez [16] | YES | 0.06 | $6 \times 10^{-4}$ | No |
| This work | YES | 0.15 | $2.4 \times 10^{-3}$ | YES |
| Amitsuka[14] | NO | 0 | 0 | NO |

Thus our work is the first observation in $URu_2Si_2$ of the clear existence of two features, closely situated, which are observed simultaneous to a ferromagnetic signature at 35 K. The presence of this ferromagnetic signature itself is probably a result of stacking fault defects and this has been noted by several workers. Weak internal fields, whose magnitude has been estimated to be anywhere between 0.1 to 10 gauss have been known to arise just below $T_{HO}$ through several NMR and µSR studies[17]. In Table I we summarize the results presented by various workers - the existence or absence of the feature at 16 K appears to depend on the degree of saturation magnetization arising from the ferromagnetic feature at 35 K.  The low temperature feature is most visible when the samples have the largest values of the saturated moment due to the ferromagnetism at 35 K. We suggest that the presence of the signature at 16 K i.e. the splitting near $T_{HO}$ is a result of the ferromagnetism arising at 35 K.

In addition to the association of the splitting near $T_{HO}$ to the ferromagnetism at 35 K our work has presented for the first time higher order AC susceptibility measurements.  Our experimental observation that $\chi_5$ is non zero while $\chi_3$ is zero could find an explanation in terms of various proposals for the existence of a spin nematic phase[18] in $URu_2Si_2$.

In summary we have reported here a new feature in DC as well as the AC susceptibility measurements in the vicinity of the hidden order transition in $URu_2Si_2$.  A peak in the first order AC susceptibility is observed at T=16 K in close proximity to the change in slope well known at the hidden order transition.  At this new transition a ferromagnetic signature is also observed in DC measurements.  This ferromagnetic signature is quickly suppressed in a small field (B> 40 Gauss ||c-axis).   We venture to suggest that the observed feature in close proximity to the hidden order transition maybe be due to the ferromagnetic "symmetry breaking field" (arising at T=35 K) which splits the complex hidden order-parameter, at least locally.    Such a scenario is not unreasonable in the context of several of the proposed models for the hidden order in $URu_2Si_2$.

**ACKNOWLEDGEMENTS:** We acknowledge useful correspondence with Piers Coleman, Pradeep Kumar, S. Ramakrishnan and Sindhunil Roy and Jim Sauls.  The work at the University of Virginia was supported by the National Science Foundation through grant DMR 0073456.  The work at Argonne National Labs was supported by the U. S. Department of Energy, Office of Science, Office of Basic Energy Sciences under Contract No. DE-AC02-06CH11357.